\def \SAIT #1 #2 {{\em Mem.\ Soc.\ Astron.\ It.\/} {\bf #1}, #2}
\def \MESS #1 #2 {{\em The Messenger\/} {\bf #1}, #2}
\def \ASTRNACH #1 #2 {{\em Astron. Nach.\/} {\bf #1}, #2}
\def \AAP #1 #2 {{\em Astron. Astrophys.\/} {\bf #1}, #2}
\def \AAL #1 #2 {{\em Astron. Astrophys. Lett.\/} {\bf #1}, L#2}
\def \AAR #1 #2 {{\em Astron. Astrophys. Rev.\/} {\bf #1}, #2}
\def \AAS #1 #2 {{\em Astron. Astrophys. Suppl. Ser.\/} {\bf #1}, #2}
\def \AJ #1 #2 {{\em Astron. J.\/} {\bf #1}, #2}
\def \ANNREV #1 #2 {{\em Ann. Rev. Astron. Astrophys.\/} {\bf #1}, #2}
\def \APJ #1 #2 {{\em Astrophys. J.\/} {\bf #1}, #2}
\def \APJL #1 #2 {{\em Astrophys. J. Lett.\/} {\bf #1}, L#2}
\def \APJS #1 #2 {{\em Astrophys. J. Suppl.\/} {\bf #1}, #2}
\def \APSS #1 #2 {{\em Astrophys. Space Sci.\/} {\bf #1}, #2}
\def \ASR #1 #2 {{\em Adv. Space Res.\/} {\bf #1}, #2}
\def \BAIC #1 #2 {{\em Bull. Astron. Inst. Czechosl.\/} {\bf #1}, #2}
\def \JSQRT #1 #2 {{\em J. Quant. Spectrosc. Radiat. Transfer\/} {\bf #1}, #2}
\def \MN #1 #2 {{\em Mon. Not. R. Astr. Soc.\/} {\bf #1}, #2}
\def \MEM #1 #2 {{\em Mem. R. Astr. Soc.\/} {\bf #1}, #2}
\def \PLR #1 #2 {{\em Phys. Lett. Rev.\/} {\bf #1}, #2}
\def \PASJ #1 #2 {{\em Publ. Astron. Soc. Japan\/} {\bf #1}, #2}
\def \PASP #1 #2 {{\em Publ. Astr. Soc. Pacific\/} {\bf #1}, #2}
\def \NAT #1 #2 {{\em Nature\/} {\bf #1}, #2}

\documentstyle{memsait}
\input epsf.sty
\begin{opening}
\title{POPULATION GRADIENTS IN NEARBY DWARF GALAXIES} 
\author{Ivo Saviane$^{1,2}$, Enrico V. Held$^3$, Yazan Momany$^{1,3}$, 
Luca Rizzi$^{3,1}$} 
\institute{
$^1$ Dipartimento di Astronomia, Universit\`a di Padova, Italy\\
$^2$ Department of Physics and Astronomy, UCLA, USA\\
$^3$ Osservatorio Astronomico di Padova, Italy}
\date{} 
\end{opening}

\providecommand{\LyX}{L\kern-.1667em\lower.25em\hbox{Y}\kern-.125emX\@}
\newcommand{\noun}[1]{\textsc{#1}}

\begin{document}

\oddpagefooter{}{}{} 
\evenpagefooter{}{}{} 
\ 
\bigskip

\begin{abstract}

Recent results on the detection of substructure in dwarf spheroidal
galaxies are discussed. In most cases they show that, when a galaxy
experiences multiple SF episodes, the intermediate age population is
more centrally concentrated than the old population, and that the
recent SF is even more concentrated towards the central regions.
Moreover, it appears that the spatial distribution of stars becomes
more and more irregular as younger and younger subpopulations are
considered.
We illustrate how wide-area observations, allowing a substantial
coverage of dwarf spheroidals (in particular the satellites of the
Milky Way), has provided evidence for cosmologically old populations
in every dSph galaxy.
All these results are placed in the broader scenario of population
gradients in nearby dwarf galaxies of different morphological types.

\end{abstract}

\section{Introduction}

The study of nearby dwarf galaxies has greatly benefited from the
progress in both space- and ground-based observational astronomy. 
HST imaging can reach the old main sequence turnoff (TO) of objects in
the Local Group (LG) even to the distance of Leo systems (Mighell \&
Rich 1996; Caputo et al. 1998; Gallart et al. 1999), and spectrographs
attached to 10m-class telescopes now permit a study of the metal
content of giant stars out to the M31 system (C\^ot\'e et al. 1999).

Another major advance has been the introduction of large-format CCDs
and mosaic cameras, which allow a substantial coverage of dwarf
spheroidal (dSph) satellites of the Milky Way.  Populations of
different age are often located in different galactic regions, so that
a spatially biased stellar sample can turn out a biased sample tout
court. It is therefore of interest to gain a broad picture of how
population gradients manifest themselves in dwarf galaxies.

\section{Population gradients in dwarf galaxies: core-halo structures}

In most cases, observations show that a centrally concentrated
population of a given age is superposed onto a more extended older
population (see below). This is a relatively recent result, since
galaxies of different morphological types present different
observational challenges.

\noindent

Asymmetries in the spatial distribution of stars in dSph galaxies have
long been known, the most striking example being that of Fornax (Hodge
1961). Centrally concentrated blue stars in the dwarf elliptical
galaxies NGC~185 and NGC~205 were known since Hodge (1963), Hodge
(1973), and Gallagher \& Hunter (1981).  Similar gradients are
observed in dwarf spheroidals, seemingly the continuation of the dE
sequence at fainter luminosities (e.g. Binggeli 1994).  Da Costa et
al. (1996) recognized a change in the HB morphology with radius in
And~I. The authors found that red, centrally concentrated, HB stars
should be \(\sim 3\div 4 \) Gyr younger than blue HB stars, and they
reached a similar conclusion for Leo~II and Sculptor by re-analyzing
existing data.

The most impressive example of population changes with radius is
perhaps the Stetson's et al. (1998) two-color investigation of
Fornax. The oldest population defines an extended halo with a spatial
scale length larger than that of intermediate-age red clump stars (\(
\sim 4\div 6 \) Gyr). The young population of blue main sequence (MS)
stars (\( \sim 100 \) Myr), as well the reddest AGB (carbon) stars,
are even more concentrated in a bar-like distribution.  The
intermediate age population displays an asymmetrical structure, with a
peculiar ``crescent'' shape. According to Buonanno et al. (1999), the
central Fornax globular cluster \#4 is
\( \sim 3 \)~Gyr younger than the rest of the clusters, which implies
that it was formed in the gap between the first SF episode and the
major intermediate-age one. We have been able to add to this picture
with wide-area \( BVI \) imaging (Saviane et al. 2000).
The RGB color distribution has a main component with a mean
metallicity {[}Fe/H{]}\( =-1.00\pm 0.15 \) and a low dispersion \(
\sigma (\rm [Fe/H])=0.12\pm 0.02 \), and there is a secondary bluer
component.  In principle this could be young and metal-rich, but its
radial distribution closely follows that of old-HB stars, so it is
unambiguously old and metal-poor.  A spatial analysis is then able to
break the age-metallicity degeneracy.
A similar degree of substructure was found in Phoenix (Held et
al. 1999). The youngest blue stars (\( 1\div 2.5\times 10^{8} \)~yr)
form clumps near the galaxy center, and are elongated in a direction
perpendicular to the major axis defined by the diffuse galaxy light
(Mart\'{\i}nez-Delgado et al. 1999a). They are slightly offset towards
the H\noun{i} cloud A observed by Young \& Lo
(1997). Mart\'{\i}nez-Delgado et al. find that the area-normalized
star formation rate for the central region of Phoenix is in the range
obtained by Hunter \& Gallagher (1986) for a sample of dIrr
galaxies. Older and intermediate age stars are spread over a larger
area. In close correspondence to Fornax, we also found 
that  \(\sim 40\% \) 
of the Phoenix red giants are of intermediate age, and more centrally
concentrated than blue HB stars. Phoenix shows a metallicity
dispersion (\( \sigma _{\rm [Fe/H]}=0.23\pm 0.03 \)) around its mean
{[}Fe/H{]}\( =-1.81\pm 0.10 \), but the RGB sample is not large enough
to allow an analysis such as that done for Fornax. 
An analysis of WFPC2 observations (Holtzman et al. 2000) leads to an
initial SF period lasting until $t\sim 9$ Gyr ago. Intermediate-age
stars are then formed ($t \sim 5$ Gyr), mildly separated in age from a
strong recent burst ($t \sim 300$ Myr).

The list of dSph/dE showing population gradients is now fairly
extended.  A centrally concentrated intermediate-age population is
found in NGC~147 (Han et al. 1997) and ESO~410-G005 (Karachentsev et
al. 2000).
In NGC~185, recent star formation (up to \( 100 \) Myr) is only
detected in the central
\( \sim 0.2\times 0.1 \) kpc\( ^{2} \) (Mart\'{\i}nez-Delgado 
et al. 1999b).  SFRs during the formation of the intermediate-age and
old populations were lower in the outer regions.
Red HB stars are more centrally concentrated than blue HB
stars in Sculptor (Hurley-Keller et al. 1999; Majewski et al. 1999)
and Sagittarius (Bellazzini et al. 1999).  In both cases, the
interpretation is made controversial (age vs. metallicity) by the yet
unsolved `second parameter' problem. 
Some counterexamples do exist where population gradients were not
detected.  Among the best studied LG dwarf spheroidals, population
gradients could not be detected in Leo~I (Held et al. 2000a; at least
outside \( 0.16 \) kpc), Carina (Da Costa et al. 1996),
Tucana (Saviane et al. 1996), and And~II (Da Costa et al. 2000).

Dwarf irregular (dIrr) and blue compact dwarf (BCD) galaxies are
generally located far from the Galaxy or the LG, yet population
gradients are readily recognized in optical and IR surface photometry
investigations. An almost ubiquitous presence of a halo of red stars
surrounding the star forming regions has been established.  As an
example, Patterson \& Thuan (1996) show that the majority of galaxies
in their dIrr sample is composed of objects having nuclear
star-forming regions superimposed to an underlying exponential
component, whose color is compatible with that of a K0 main-sequence
star or G5 giant. K and M giants are also the probable old population
in the BCD and dIrr samples of Thuan (1983, 1985). Other examples of
extended halos in BCDs are found in Loose \& Thuan (1986), Kunth et
al. (1988), Meurer et al. (1994), Papaderos et al. (1996), Telles \&
Terlevich (1997), Doublier et al. (1997; 2000), Marlowe et al. (1999).
Interestingly, Doublier et al. (2000) revealed an underlying
(unresolved) red population in POX 186, the prototype of a subclass of
BCDs that seem to have no extended halos.  Kunth and \"Ostlin (2000)
have recently reviewed this subject.

\section{The age of the extended halos: ubiquitous old populations?}

How old are these halos? This is an important question, since a
definite proof that any dwarf galaxy contains a truly old population
(i.e. \emph{at least} as old as Milky Way halo globulars), would tell
us that star formation began at the very same time throughout the
Universe, and that dwarf galaxies, irrespective of their present
morphology, are truly primordial objects. According to current stellar
evolution theories, the only unambiguous signature of a metal-poor, \(
>10 \)~Gyr population is the presence in color-magnitude 
diagrams (CMDs) of an
\emph{extended} horizontal branch (HB). Indeed, even if the TO is
reached, the uncertainties on the distance can hamper the age-dating
process, since a typical \( 0.1 \) error in distance modulus (rising
often to \( 0.2 \) magnitudes), translates into a \( \sim 1\div 2 \)
Gyr age uncertainty. 
The HB can be betrayed by the presence of RR Lyr variables,
but in order to constrain their age range, a detailed study of their
properties (e.g. metallicity, period distribution, period-amplitude
relation, etc.) should be done. In fact, RR Lyr can exist for ages
significantly younger than Milky Way 
halo GCs (e.g. Olszewski et al. 1987).

The earliest detections of a blue HB in dSph are those of Draco (Baade
\& Swope 1961) and  Ursa Minor (van Agt 1967), so there is no question 
about the fact that at least some spheroidals do contain very old stars.
Some clues on the presence of an old stellar population in 
\emph{all} of the Local Group dwarf spheroidals came by recent results. 
Smith et al. (1998) and Bellazzini et al. (1999) provided evidence for
an old/metal-poor population in the Sagittarius dwarf, and a wide-area
search with the ESO New Technology Telescope led our group to the
discovery of a significant old population in the Leo~I dSph (Held et
al. 2000a).  Wide-field ESO~2.2m data are being analyzed with the aim
of discovering and studying the Leo~I RR Lyr population (Clementini et
al. 2001, in preparation).

A special case is represented by a class of dwarf galaxies with
characteristics intermediate between those of dwarf irregulars and
dwarf spheroidals. Mateo (1998) lists 5 galaxies in this class:
Phoenix, LGS3, Antlia, DDO~210 and Pegasus.
Whether or not these are dIrr in the process of transforming into
dSph, they are expected to contain an old population.
Indeed, this is probably the case.  A blue HB was revealed in Phoenix
by Held et al. (1999), by selecting stars in the outer regions
(outside \( 0.16 \)~kpc from the center), while detection of a large
number of candidate RR Lyrae variables has been reported by Held et
al. (2000b). An old population has also been detected in DDO~210
(Tolstoy et al. 2000).

For star forming dwarf galaxies, the identification of the extended HB
is more difficult, since it can be masked by an intermediate-age 
subgiant branch of \( \sim 1 \) Gyr (depending on the metallicity) 
or by the blue MS of the youngest populations.  
HST studies of the resolved populations in BCD's
 have been carried out for, among others, VII~ZW~403
(Schulte-Ladbeck et al. 1999), I~ZW~18 (Aloisi et al. 1999;
\"Ostlin 2000), UGC~6456 (Lynds et al. 1998) and SBS~1415+437 (Thuan
et al. 1999). The first three all show populations of giants
compatible with an age of several Gyr, but in all cases the HB 
is beyond the reach of the HST.  
Further circumstantial evidence of the existence of a cosmologically 
old population in BCDs is the existence of a system of globular clusters
around Mrk~996 (Thuan et al. 1996), whose integrated colors are
similar to those of NGC~6752 or \( \omega \)Cen.

Evidence for old populations is stronger in dwarf irregulars. 
RR Lyrae variables
were detected in the field of IC~1613 (Saha et al. 1992), and
there are also some hints of a blue HB in the CMD of its central
regions (Cole et al. 1999), but it is confused among young main
sequence  stars. The HB is visible in the CMD of WLM presented in
Rejkuba et al. (2000), for a region between disk and halo 
imaged with HST/STIS. A blue HB was also revealed in the WLM globular
cluster (Hodge et al. 1999).

To summarize all these results, it appears that, when careful
investigations are carried out, old populations invariably come out
even in galaxies clearly dominated by young or intermediate age stars.

\section{Summary}

A real understanding of the evolution of nearby dwarfs requires data
samples that are both deep and provide full coverage of the galaxy
extension. When such samples are available, cosmologically old
populations almost invariably turn up. It therefore appears that there
has been a generalized primordial SF episode in any dwarf galaxy
(besides the yet undetected  population III).
Successive evolution is much varied and not fully understood, although
initial gas density, angular momentum and later mutual interactions
all may have a role.
It also appears that, when SF is not prematurely halted (e.g. by SNII
explosions and removal of the ISM), a major intermediate age episode
is often experienced by dSph. Later SF is also observed in some dSph,
and recent SF becomes active or very active in dIrr and BCDs. The
luminosity-weighted SFH for five Galactic dSph of Hernandez et
al. (2000) seems to support this scenario.

In most cases galaxy evolution leads to more centrally concentrated
younger populations, so apparently the gas contracts while forming new
stars. Moreover, there is evidence that the central populations are
also more metal-rich (for example in NGC~147: Han et al. 1997), so it
is plausible that the gas becomes enriched by each SF burst.
Small remaining amounts of gas are observed in some objects, that
could have been released by the last
generation of stars, and either retained (e.g. NGC~185 --
Mart\'{\i}nez-Delgado et al. 1999b) or blown out (e.g. Phoenix -- 
Young \& Lo 1997).
The spatial distribution of
stars  becomes more and more irregular as
younger and younger subpopulations are considered, a fact that
resembles what is observed in more active dwarf irregular galaxies.

\end{document}